\documentclass[aps,prd,twocolumn,superscriptaddress,preprintnumbers,floatfix,nofootinbib,notitlepage,showkeys,showpacs]{revtex4-2}

\usepackage[svgnames]{xcolor}
\usepackage{amsthm,bbm}
\usepackage{float}

\usepackage{siunitx}

\usepackage{tikz}
\usetikzlibrary{decorations.markings, arrows}

\usepackage{graphicx,times}
\usepackage{latexsym}
\usepackage{mathtools}

\usepackage{amsmath,amssymb,amsbsy,amsfonts}
\usepackage{array}
\usepackage{bm}
\usepackage{mathrsfs}
\usepackage{cancel}
\usepackage[normalem]{ulem}

\usepackage{caption}
\usepackage{ragged2e}
\DeclareCaptionJustification{justified}{\justifying}
\usepackage{subfig}
\usepackage{setspace}
\usepackage{xcolor}
\usepackage{makecell}
\newcommand\TopRule{\Xhline{0.08em}}

\newcommand\MidRule{\Xhline{0.03em}}
\newcommand\BotRule{\Xhline{0.08em}}

\newcommand{\phiv}{{\pmb{\phi}}}

\renewcommand\>{\rangle}

\usepackage[hyperfootnotes=true,colorlinks]{hyperref}
\hypersetup{
    colorlinks,
    allcolors=blue!50!black,
  }
\usepackage[capitalise]{cleveref}

\begin{document}

\title{Space-time symmetric qubit regularization of the asymptotically free two-dimensional O(4) model}
\preprint{LA-UR-21-30982, INT-PUB-21-023, IQuS@UW-21-013}
\author{Junzhe Zhou}
\email{junzhe.zhou@duke.edu}
\affiliation{Department of Physics, Box 90305, Duke University, Durham, North Carolina 27708-0305, USA}
\author{Hersh Singh}
\email{hershsg@uw.edu}
\affiliation{
InQubator for Quantum Simulation (IQuS), Department of Physics, University of Washington, Seattle, Washington 98195-1550, USA}
\author{Tanmoy Bhattacharya}
\email{tanmoy@lanl.gov}
\affiliation{Los Alamos National Laboratory, Los Alamos, New Mexico 87545, USA}
\author{Shailesh Chandrasekharan}
\email{sch27@duke.edu}
\affiliation{Department of Physics, Box 90305, Duke University, Durham, North Carolina 27708-0305, USA}
\author{Rajan Gupta}
\email{rg@lanl.gov}
\affiliation{Los Alamos National Laboratory, Los Alamos, New Mexico 87545, USA}

\begin{abstract}
  We explore if space-time symmetric lattice field theory models with a finite Hilbert space per lattice site can reproduce asymptotic freedom in the two-dimensional $O(4)$ model. We focus on a simple class of such models with a five dimensional local Hilbert space. We demonstrate how even the simplest model reproduces asymptotic freedom within the D-theory formalism but at the cost of increasing the size of the Hilbert space through coupling several layers of a two-dimensional lattice. We then argue that qubit regularization can be viewed as an effective field theory (EFT) even if the continuum limit cannot be reached, as long as we can tune the model close enough to the continuum limit where perturbation theory, or other analytical techniques, become viable. We construct a simple lattice model on a single layer with a four dimensional local Hilbert space that acts like an excellent EFT of the original theory.
\end{abstract}

\maketitle

\section{Introduction}

Static properties of strongly coupled relativistic quantum field theories (QFTs) can usually be obtained non-perturbatively using a ``traditional'' space-time symmetric Euclidean lattice field theory \cite{Wilson:1974sk,Kogut:1979wt}. The lattice acts as an ultraviolet regulator and continuum QFTs are obtained by approaching the critical points of the lattice theory. In QFTs with microscopic bosonic degrees of freedom, the local Hilbert space of these traditional lattice models are always infinite dimensional. An interesting theoretical question, motivated by quantum computation of QFTs, is whether there are space-time symmetric lattice models with a finite dimensional local Hilbert space that can also reproduce a desired quantum field theory. Such a finite dimensional approach to QFTs, especially in the Hamiltonian formulation, has become an active area of research due to the possibility of studying QFTs using a quantum computer~\cite{Jordan:2011ne,Casanova:2011wh,Casanova:2012zz,PhysRevX.6.031007,Klco:2018kyo,Klco:2019evd,Alexandru:2019nsa,Banuls:2019bmf,Banuls:2019rao,Alexandru:2019ozf,Raychowdhury:2019iki,Bender:2020jgr,Alexandru:2021xkf}. It can be viewed as a new type of regularization of QFTs called  ``qubit regularization'' \cite{Singh:2019uwd}.\looseness-1

When qubit regularization of a Lorentz-invariant field theory is constructed for carrying out quantum evolution, it seems natural to consider the Hamiltonian formulation, in which  the manifest symmetry between space and time is lost. One expects that in the Hamiltonian approach, the space-time symmetry emerges dynamically at the quantum critical point. Whether this occurs or not is, in general, a subtle question. The existence of an explicitly space-time symmetric Euclidean approach could alleviate such concerns, and it would be strange if restricting to a finite-dimensional local Hilbert space allows only a Hamiltonian approach to a relativistic field theory.
Furthermore, relativistic QFTs are often \emph{defined} as an analytic continuation of the well understood Euclidean QFTs from imaginary to real time.
In a space-time symmetric approach, then, the Euclidean transfer matrix may motivate quantum circuits for the discretized time-evolution, without recourse to the usual Trotterization schemes.\looseness-1

Motivated by these points, we look for space-time symmetric qubit regularized asymptotically free field theories. We already know of several examples of such regularizations of QFTs that emerge at infrared fixed points. The most well known of these is the Ising model, which is a space-time symmetric lattice model with a two-dimensional local Hilbert space on each lattice site, but can reproduce the $Z_2$ Wilson-Fisher fixed point in three dimensions and the Gaussian fixed point in four dimensions. In the continuum, this fixed point is naturally formulated with an infinite-dimensional local Hilbert space of a single component real scalar field. Similarly, the physics of the BKT transition that arises in a two-dimensional\footnote{By abuse of terminology, we refer to both a \(1+1\)-dimensional Minkowski QFT and the corresponding two-dimensional Euclidean theory as two-dimensional theories.} field theory with a single complex scalar field can be reproduced using various space-time symmetric qubit regularizations \cite{PhysRevB.103.245137}. Recently, $O(N)$ symmetric Wilson-Fisher and Gaussian fixed points have been reproduced with space-time symmetric loop models with
finite dimensional Hilbert spaces \cite{Singh:2019uwd,Singh:2019jog}. Hamiltonian approaches to these fixed points in qubit regularized models are well known in condensed matter physics, for example, the $O(3)$ Wilson-Fisher fixed point was studied in \cite{PhysRevLett.72.2777}.

Does a space-time symmetric qubit regularization, i.e., a theory with a finite local Hilbert space, exist for asymptotically free QFTs? Indeed it does within the D-theory approach \cite{Brower:2003vy,Wiese:2006kp}. In this approach, an extra spatial dimension is introduced, which plays the role of the inverse coupling of the theory. As the size of this dimension increases, the physical correlation length increases exponentially. This suggests that even asymptotic freedom is recovered exponentially fast in the size of the extra dimension. From the perspective of qubit regularization, adding an extra dimension can be viewed as increasing the size of the local Hilbert space. However, both the exponentially fast approach to criticality and the locality of interactions in this expanded space makes this an attractive solution both in the Hamiltonian and Euclidean formulations. Several earlier results already show this feature of D-theory using either a space-time asymmetric Hamiltonian formulation \cite{Evans:2018njs}, or by using an extra dimension which could be either continuous \cite{Bietenholz:2003wa, Beard:2004jr} or discrete. Our goal here is to demonstrate that D-theory also reproduces asymptotic freedom if we approach it as several coupled layers of a space-time symmetric qubit-regularized model.  \looseness-1

We further argue that even when one cannot reach the continuum limit with a finite number of extra layers, the resulting constructions can be thought of as useful effective field theories for the target theory, provided the physics above the cutoff can be described by perturbation theory or other controlled techniques. In this paper, we show that by introducing and tuning an additional coupling even in the single layer model one can reach such large correlation lengths. Alternately, if there exists an asymptotically-free critical point in a single-layer space-time symmetric model then the need for a D-theory approach is unnecessary unless it can be simulated with fewer quantum resources. The search for such a critical point is still ongoing. 

In this work, we focus on recovering the asymptotic freedom of two-dimensional $O(4)$ model using a finite dimensional Hilbert space. Since an $O(4)$ model can be viewed as as $SU(2)$ matrix model, our qubit models may be extendable to qubit models of $SU(2)$ gauge theories through simple modifications.\looseness-1

This paper is organized as follows. In \cref{sec:2}, we compute the step scaling function (SSF) of the traditional $O(4)$ model defined on the infinite dimensional local Hilbert space. We then explore if our qubit models can recover this SSF. In \cref{sec:3}, we show how a simple model on a five dimensional Hilbert space is able to reproduce the SSF of the $O(4)$ model through the D-theory approach. In \cref{sec:4}, we argue that, in the absence of a critical point, we may be able to view qubit regularization as an effective field theory. We present our conclusions in \cref{sec:5}.

\section{Step Scaling Function}
\label{sec:2}

A reliable way to quantitatively understand if a qubit model reproduces the asymptotically free QFT of interest, is to compute the step scaling function $\sigma(s,u)$ that characterizes it \cite{Luscher:1991wu}. This function is defined in the continuum through a physical dimensionless coupling $u(L)$ defined in a finite  box of size $L$ and describes how it changes when the size of the box is scaled by a factor $s$. More precisely, we define 
\begin{align}
\sigma(s,u(L)) = u(sL).
\end{align}
The definition of the dimensionless coupling $u(L)$ is not unique. One definition is given by $u(L) = m(L) L$ where $m(L)$ is the finite size mass gap \cite{Luscher:1991wu}.
In this work we will, instead, use $u(L) = L/\xi(L)$, where $\xi(L)$ is the correlation length.  
We choose to use the second-moment definition of $\xi$ introduced long ago \cite{Cooper:1982nn,PhysRevB.45.2883}. 

\begin{figure}[htb]
\includegraphics[width=\hsize]{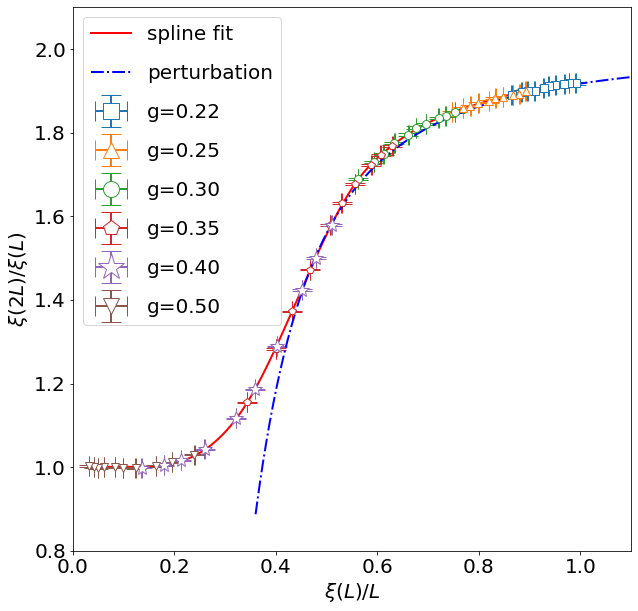}
\caption{The step scaling function of the traditional $O(4)$ nonlinear sigma model, \cref{eq:o4-trad}, determined using the Wolff cluster algorithm with error bars denoting $1\sigma$ confidence intervals. A cubic smoothing-spline is used to piece together the data simulated at a number of values of $g$ to obtain a universal step scaling function which is believed to characterize the $O(4)$ model. This is used to benchmark our calculations of the qubit-regularized $O(4)$ models.}
\label{fig:ssftrad}
\end{figure}

Since the focus of our work is the asymptotically free two-dimensional $O(4)$ QFT, we first compute the step scaling function of this model using the traditional lattice $O(4)$ model, formulated in terms of real unit four-vectors $\phiv_x$ at every lattice site $x$. The Euclidean action of this model is given by
\begin{align}
S = -\frac{1}{g} \sum_{\langle x y\rangle}   \phiv_x\cdot \phiv_y,
\label{eq:o4-trad}
\end{align}
where \(g\) is the coupling and the sum is over all bonds $\<xy\>$ between nearest-neighbor sites $x$ and $y$. This model was studied in Ref.~\onlinecite{Caracciolo:1992nh} from an algorithmic perspective; here we repeat the calculations using the Wolff cluster algorithm \cite{Wolff:1988uh,Wolff:1989hv} and compute $(\xi(2L)/\xi(L) = 1/(\sigma(2,u))$ as a function of $\xi(L)/L = 1/u$. In order to compute $\xi(L)$ using the second-moment definition, we first compute the momentum-space two-point correlation function 
\begin{align}
\tilde{G}(p) =  \frac{1}{ZL^2} \sum_{x,y} e^{ip \cdot (x-y)} \int [d\phiv] \ e^{-S} \phiv_x \cdot \phiv_y
\label{eq:corrfn}
\end{align}
where $\phiv_x$ and $\phiv_y$ are $O(4)$ vectors at sites $x$ and $y$, $Z = \int [d\phiv] \ e^{-S}$ is the partition function, and $L$ is the size, in lattice units, of our two-dimensional square lattice. The second-moment definition of the correlation length is given by \cite{Caracciolo:1992nh},
\begin{align}
\xi(L) =  \frac{1}{2\sin(\pi/L)}\sqrt{\frac{\chi}{F}-1}
\label{eq:corrlen}
\end{align}
where $\chi = \tilde{G}(p)|_{|p|=0}$ and $F = \tilde{G}(p)|_{|p|=2\pi/L}$.

Our result for the step scaling function for the $O(4)$ model is shown in \cref{fig:ssftrad}. The calculations were done at couplings $g= \num{0.22},\num{0.25},\num{0.30},\num{0.35},\num{0.40},$ and $\num{0.50}$ on lattices ranging from $L=32$ up to $L=512$. The data was pieced together using an univariate cubic spline with a smoothing scale of 0.0003. We did not include data with $L<32$ to avoid discretization errors.

\begin{figure*}[!htbp]
\includegraphics[width=0.45\hsize]{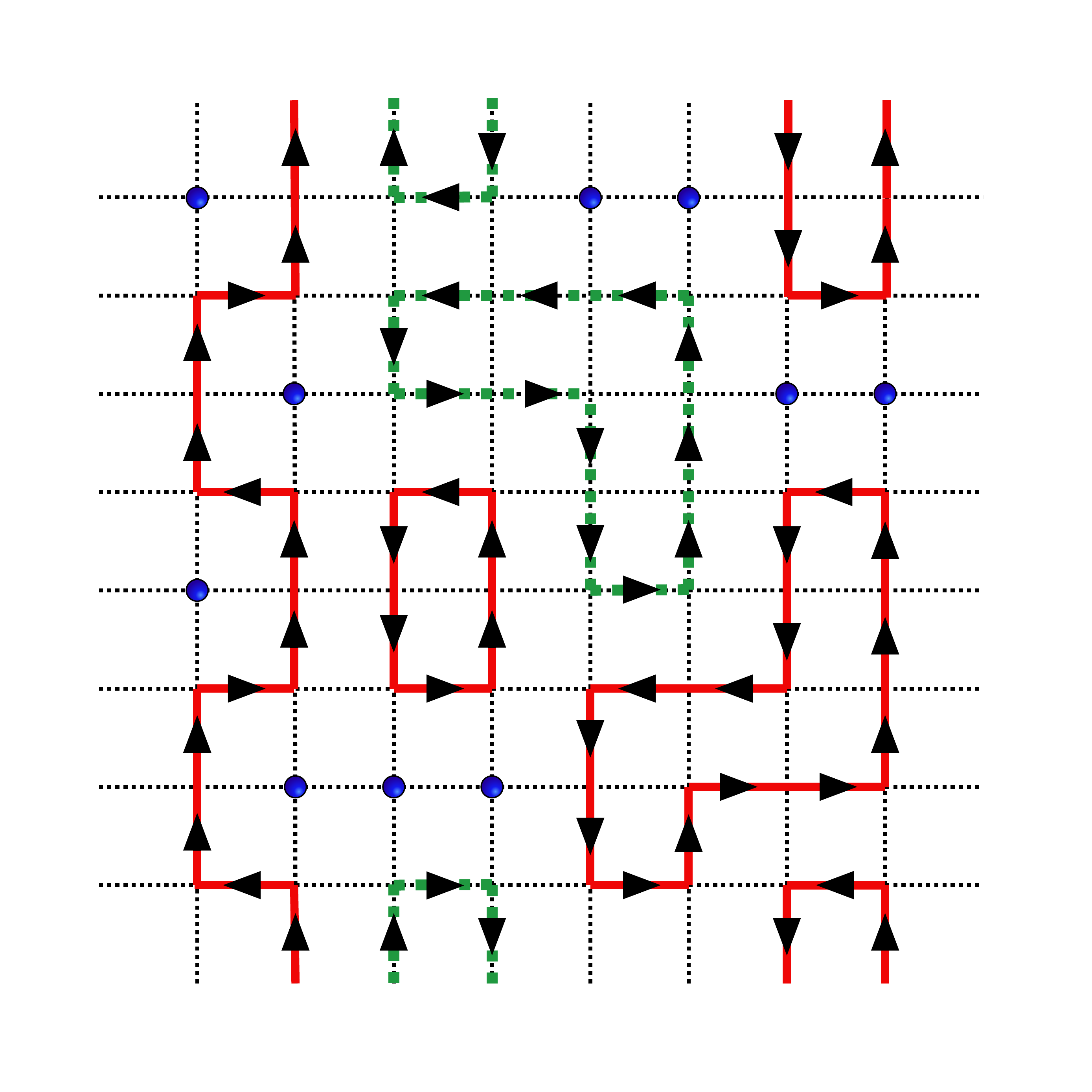}
\includegraphics[width=0.45\hsize]{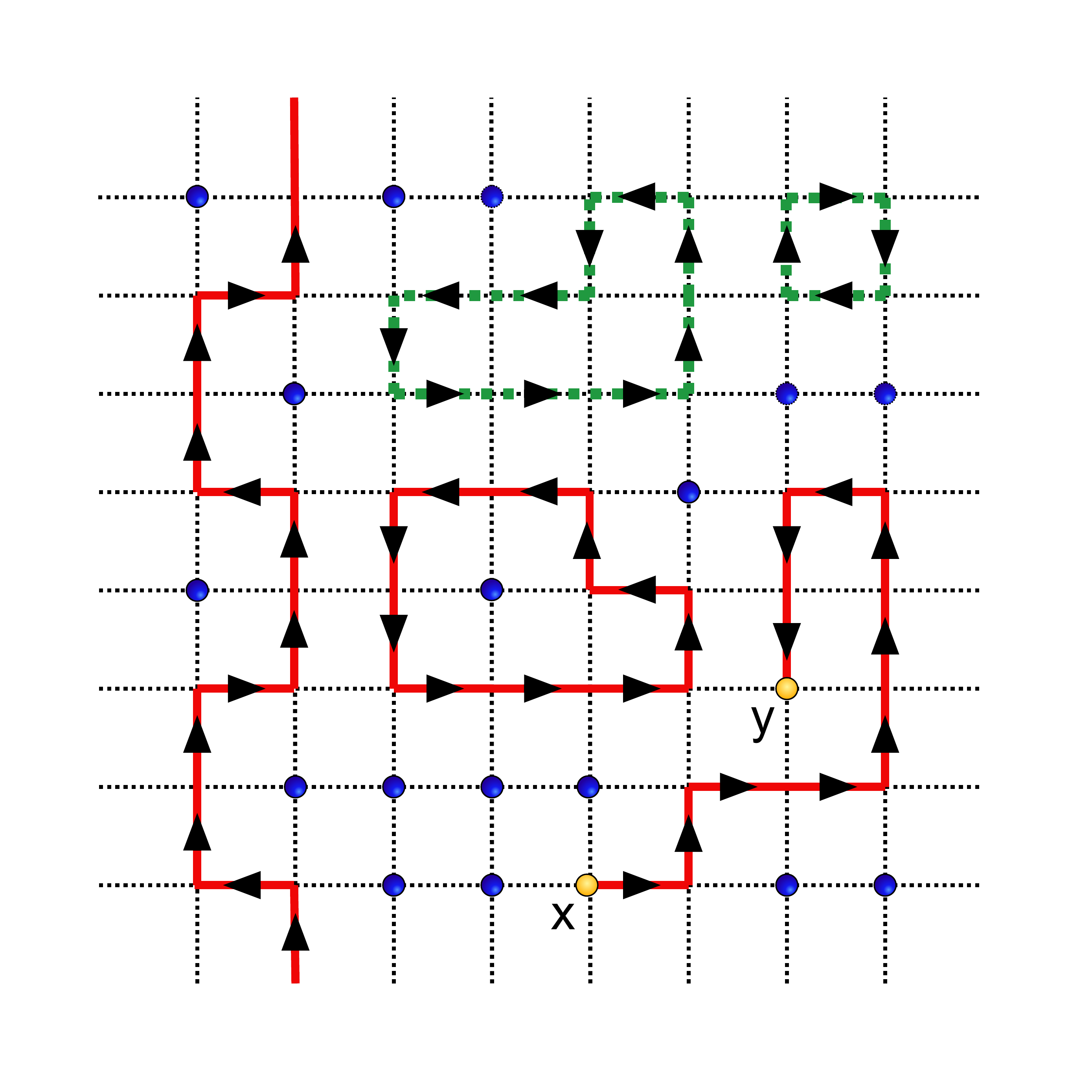}
\caption{The left figure shows a configuration of the qubit-regularized $O(4)$ lattice model whose partition function is given in \cref{eq:mod1}. It has a five-dimensional Hilbert space at each site which consists of  singlets (monomers) represented by blue circles, and worldlines of $O(4)$ vector particles represented by two flavors (red and green) of oriented loops shown as solid and dotted lines. In the D-theory extension, configurations have $L_y$ such two-dimensional layers and the particle world lines can traverse between the layers. The right figure is an illustration of a configuration ${\cal C}_{x,y}$ containing an open $O(4)$ worldline of a particular flavor with a creation and annihilation operator at lattice sites $x$ and $y$.
}
\label{fig:conf1}
\end{figure*}

\section{Qubit Regularization using D-theory}
\label{sec:3}

In this section, we focus on reproducing the step-scaling function {in \cref{fig:ssftrad}} using a qubit-regularized space-time symmetric lattice field theory with a finite Hilbert space per lattice site. In the traditional model, the $O(4)$ vector lies in an infinite dimensional Hilbert space $\mathbb{V}_{\rm trad}$, constructed as a direct sum of a specific set of irreducible representations of the $O(4)$ symmetry group. A general $O(4)$ irreducible representation (irrep) can be characterized by  left and right $SU(2)$ spin-$j$ irreps and labeled as $(j_L,j_R)$. If we label the corresponding vector space of the irreps as $\mathbb{V}_{j_L,j_R}$, then
the local Hilbert space of the traditional model can be decomposed as
\begin{align}
\mathbb{V}_{\rm trad} = \bigoplus_{j=0,1/2,1,...} \mathbb{V}_{j,j}.    
\end{align}
Simple qubit-regularized models emerge from a vector space where we {keep only} the leading two irreps,
\begin{align}
\mathbb{V}_{\rm qubit} = \bigoplus_{j=0,1/2} \mathbb{V}_{j,j}.
\end{align}
In such models, each site can be either empty (singlet, $j=0$) or filled (vector, $j = 1/2$). 
Conservation of $O(4)$ currents imply that, in the worldline representation, filled states form closed loops. One of the simplest of these models is a loop-gas model, whose partition function can be written as
\begin{align}
Z = \sum_{{\cal C}} U^{N_m[\cal C]} 
\label{eq:mod1}
\end{align}
where a configuration ${\cal C}$ consists of empty sites (monomers) and closed loops formed by connecting nearest neighbor sites (worldlines), and $N_m[\cal C]$ is the number of monomers. Each loop, corresponding to the vector representation of the $O(4)$, can be in one of  four states, which we represent by two colors and two orientations. Such space-time symmetric models in this five-dimensional Hilbert space can be motivated by strongly coupled lattice QED with staggered fermions \cite{Cecile:2007dv}: except we do not allow worldlines to hop to a nearest neighbor site and back immediately since that does not have a nice interpretation in the Hamiltonian limit. An illustration of configuration ${\cal C}$ is shown in the left panel of \cref{fig:conf1}. 

The simplicity of this simple five-dimensional Hilbert space model is striking: the only tuning parameter is the coupling $U$ that controls the number of monomers. It was not obtained by truncating the Hamiltonian of the traditional model, nor will we obtain the continuum limit by modifying the model to include more irreps.
 Even though the loop gas models are very simple, in three dimensions, they can reproduce the physics of the $O(4)$ Wilson-Fisher fixed point \cite{Banerjee:2019jpw,Singh:2019jog}.
This, of course, is not surprising since once we have a model within the right basin of attraction of the renormalization group flow, the details of the model should not matter close to second-order critical points. 

As we already discussed in the introduction, there is no reason for this simple model to be able to reproduce asymptotic freedom of the traditional two-dimensional model, especially close to the continuum limit. For a generic value of $U$, we expect the model to develop a mass gap in two dimensions, but it is not clear if there is a critical coupling $U_c$ where this mass gap vanishes. Even if such a critical point exists, it is not at all obvious that it is the asymptotically free $O(4)$ model describing physics at all scales. On the other hand, D-theory suggests that starting with couplings that put the 3-D $O(4)$ model in the broken phase, asymptotic freedom in the corresponding  two-dimensional model with $L_y$ coupled layers will improve as $L_y$ is increased \cite{Brower:2003vy}. Here, we show results that confirm this prediction of D-theory.

\begin{figure*}[!htbp]
\centering
\includegraphics[width=0.45\hsize]{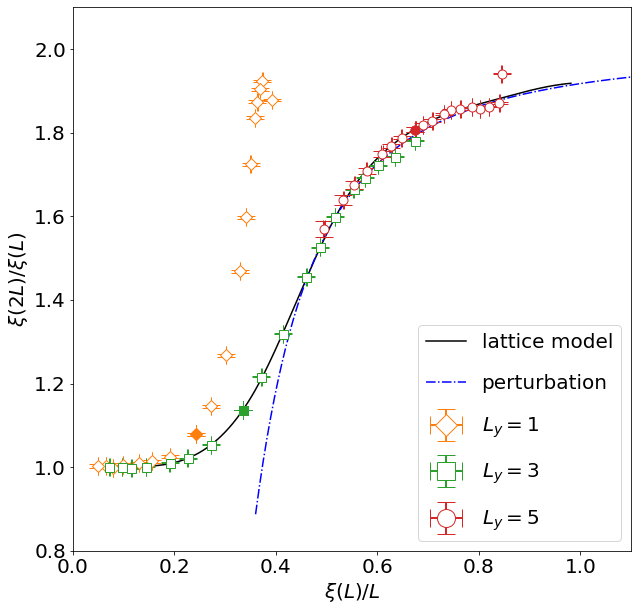} 
\includegraphics[width=0.45\hsize]{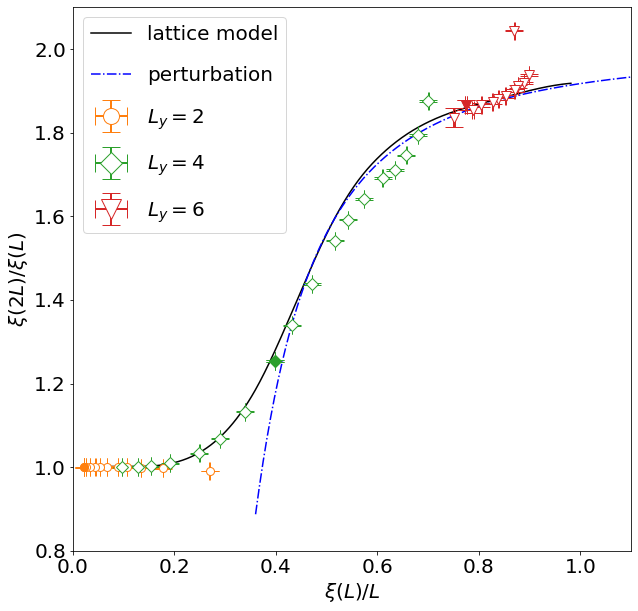}
\caption{Step scaling functions of the qubit model described by \cref{eq:mod1} with $U=0.1$. Results for $L_y=1,3,5$ are shown in the left panel, and for $L_y=2,4,6$ in the right. The universal step scaling function for the traditional $O(4)$ model obtained in \cref{sec:2} is shown as the solid line, and the result of two-loop perturbation theory from \cite{Caracciolo:1994ud} is shown as dashed lines in the two panels. Points with \(L=48\) are highlighted using filled symbols to display the dependence of \(\xi(48)\) on \(L_y\).}
\label{fig:ssf-Dtheory}
\end{figure*}

We consider the model described by \cref{eq:mod1}, but on $L_y$ layers of a square lattices of size $L$ and with periodic boundary conditions in all three dimensions.  For simplicity, we choose $U=0.1$, since at that value of the coupling the three-dimensional $O(4)$ model, obtained by approaching the thermodynamic limit with $L_y=L$, is in the broken phase. In fact, we have evidence that this three dimensional model undergoes a phase transition to the symmetric phase at $U_c \approx \num{3.85(5)}$. We find that by increasing $L_y$ at $U=\num{0.1}$, but keeping $L_y\ll L$, our model describes the asymptotically-free two-dimensional $O(4)$ model. This is shown by calculating the step scaling function $\sigma(2,u)$ as a function of $u$ and $L_y$.

We calculate $\xi$ and compute $\sigma$ as explained in \cref{sec:2}. To calculate $\xi$ using  \cref{eq:corrlen}, we define new  configurations ${\cal C}_{x,y}$ with a worldline that contribute to the two-point correlation function $\tilde{G}(p)$ in the qubit model by introducing a creation and annihilation operator placed at two sites $x$ and $y$ as shown in the right panel of \cref{fig:conf1}.  These correlation-function configurations are not part of the partition function $Z$, which is still given by \cref{eq:mod1}, but define $\tilde{G}(p)$ though the relation
\begin{align}
\tilde{G}(p) = \frac{1}{ZL^2} \sum_{{\cal C}_{x,y}} U^{N_m[\cal C]}\ e^{i p\cdot (x-y)}
\end{align}
Monte Carlo methods based on worm updates can naturally compute this momentum-space correlation function $\tilde{G}(p)$. Here $x$ and $y$ are the locations of the head and tail of the worm and, in our method, the worm starts at some random site $x$ and samples configurations $C_{x,y}$ with the right weight.
Since we fix $U=0.1$ in this study, there are no free parameters in the model except the number of layers $L_y$, which plays the role of the coupling $1/g$ in the traditional model. As $L_y$ becomes large, we expect to reach the continuum limit. We explore $L_y$ between $1$ and $6$, and for each $L_y$ compute $\xi(L)$ using \cref{eq:corrlen} with $L$ between $4$ and $512$ or until the correlation length begins to saturate. \Cref{fig:ssf-Dtheory} shows the step scaling function of each of these D-theory models. We display odd $L_y$ (left figure) and even $L_y$ (right figure) in separate panels since they seem to behave differently, at least for small values of $L_y$. For example, data in~\cref{tab:corrDth} show that $\xi({\infty})$ of the $L_y = 2$ model is roughly $1$ in lattice units while that with $L_y=1$ is already $12$. Though
we have not investigated why this is the case in our model, we wish to point out something similar occurs in quantum spin-half chains versus ladders. In that case, while the spin-half chain is critical (i.e., has infinite correlation length), the ladder has a finite correlation length. In our case, what is more important is that this difference between even and odd $L_y$ becomes small for $L_y > 2$, and the $O(4)$ universal step scaling function is reproduced at increasingly larger values of $\xi(L)/L$ as $L_y$ is increased.

\begin{table}[!htb]
\centering
\renewcommand{\arraystretch}{1.4}
\setlength{\tabcolsep}{4pt}
\begin{tabular}{c|ccccc}
\TopRule
$L_y$ & $L_a$ & $\xi$($L_a$) & $\xi(L_a)/L_a$ & $\xi_\infty$ & $\xi\ (L_a)/\xi_\infty$ \\ 
\MidRule
1 & 96 & 12.52(2) & 0.13 & 12.66(3) & 0.99 \\
2 & 12 & 1.066(1) & 0.09 & 1.066(1) & 1.00 \\
3 & 12 & 6.64(1) & 0.55 & 18.45(5) & 0.36 \\
4 & 64 & 21.67(2) & 0.34 & 24.76(2) & 0.87 \\
5 & 16 & 12.22(1) & 0.76 & 250(10) & 0.048 \\
6 & 16 & 13.65(2) & 0.85 & 860(30) & 0.015 \\
\BotRule
\end{tabular}
\caption{\label{tab:corrDth} Correlation lengths reached in the D-theory approach at $U=0.1$ as we vary the number of layers $L_y$. We also give $L_a$  (the smallest lattice size where the model begins to agree with the universal step scaling function) and the corresponding $\xi(L_a)$.}
\end{table}

We observe that the step-scaling function of the qubit regularized model falls on the universal curve below a certain value of $\xi(L)/L$ that is $L_y$ dependent. For a given $L_y$, we denote by $L_a$ the value of $L$ at which this takes place. This $\xi(L_a)$ acts like an ultraviolet cutoff for our target theory in the sense that simulations with $\xi(L)/L \le \xi(L_a)/L_a$ can be used to find the universal step-scaling function. Knowing the step-scaling function, any such simulation can be used to compute the infinite volume correlation length $\xi_\infty = \xi(L\rightarrow \infty)$. For example, let $\xi(L_a)$ be the finite size correlation length at $L=L_a$, then using the step scaling function we can compute $\xi(2L_a)$ and repeat the process to obtain $\xi(2^k L_a)$ for arbitrary values of $k$. In this iterative process, the new 
$\xi(2L)/2L$ lies to the left of the previous $\xi(L)/L$ for $\xi(2L)/\xi(L) < 2$, 
i.e., one flows to the left along the curve. In \cref{tab:corrDth}, we list the values of $L_a$, $\xi(L_a)$ and $\xi_\infty$ for various values of $L_y$. A simple quantitative measure of how close a model is to the continuum limit is provided by the dimensionless ratio $\xi(L_a)/\xi_\infty$ given in the last 
column of \cref{tab:corrDth}. This ratio is
analogous  to the ratio of the mass gap to the ultraviolet cut off in a model, i.e., the mass 
gap in lattice unit, which goes to zero in the continuum limit. Similarly, $\xi(L_a)/\xi_\infty$ should approach zero as happens when $L_y$ is increased, while it is of order one for small $L_y$, i.e., the region dominated by lattice artifacts.

\begin{figure}[!htbp]
    \centering
    \includegraphics[width=0.95\hsize]{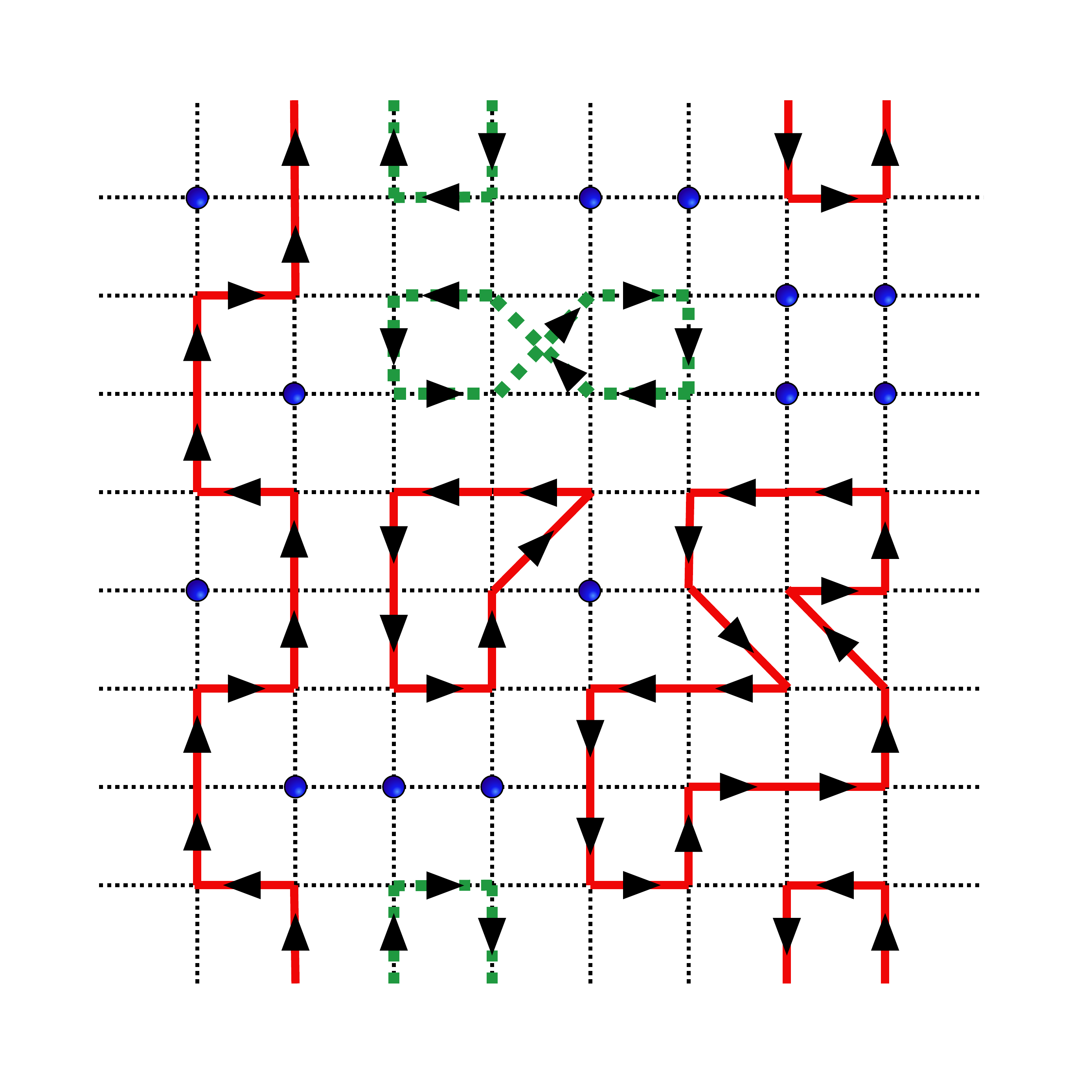}
    \caption{An illustration of the configuration of the extended qubit-regularized $O(4)$ lattice model with couplings $U$ (for vacuum sites) and $J$ (for diagonal hops). The nearest neighbor hops have weights of $1$ as before.}
    \label{fig:conf2}
\end{figure}

\begin{figure}[!htbp]
    \centering
    \includegraphics[width=0.95\hsize]{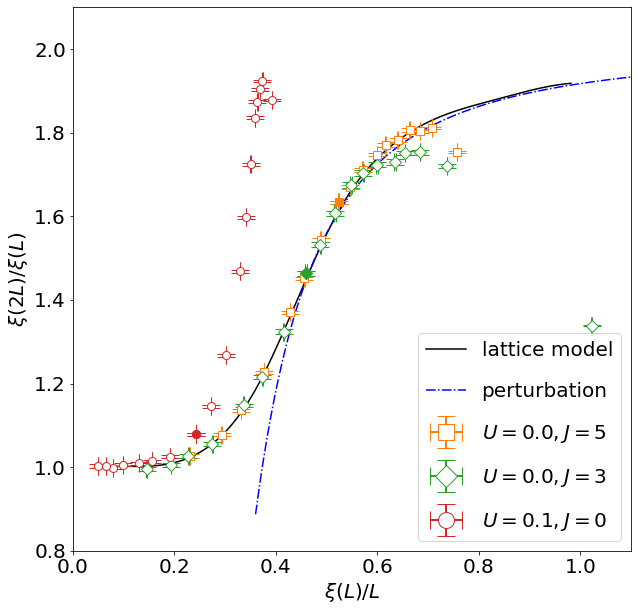}
    \caption{The step scaling function of the extended model with $U=0.1,J=0$ (which is the same as the D-theory model), $U=0,J=3$ and $U=0,J=5$. The solid line is the continuum curve, while the dashed line is the two-loop result. {Points with L=48 are highlighted with filled symbols to display the dependence of $\xi(48)$ on the parameters $U$, $J$.}}
    \label{fig:qeft}
\end{figure}

\section{Qubit regularization as an EFT}
\label{sec:4}

As we have demonstrated, D-theory is able to reproduce asymptotic freedom of the continuum two-dimensional $O(4)$ model even with a simple space-time symmetric qubit-regularized model, but at the cost of increasing $L_y$ (the number of coupled two-dimensional layers of lattices) systematically to larger values. An interesting related question is whether we can reproduce the same continuum physics with a single layer model by further exploring the model space. This was realized for the $O(3)$ model where an interesting critical Hamiltonian model was identified \cite{Bhattacharya:2020gpm}.
Discovering a space-time symmetric qubit-regularized model,
however, remains an interesting challenge for both the $O(3)$ and $O(4)$ cases.

Instead of finding a model that has a continuum limit, one can ask a simpler question. Can a lattice theory be devised which can be taken sufficiently close to the continuum limit with just a single layer by tuning the parameters, and beyond which one can extrapolate to the continuum limit using certain order perturbation theory? In that case, qubit regularization can be viewed as finding the right EFT for the purpose of quantum computation. From \cref{fig:ssftrad}, we
observe that for the $O(4)$ case, perturbation theory begins to work reasonably well when $\xi(L)/L > 0.5$. So if we can find a qubit regularization where $\xi(L_a)/L_a > 0.5$ as defined in the previous section, then it would be an excellent EFT of the $O(4)$ model. This criterion is not satisfied by the simplest model that we explored in the previous section. We, therefore, explored an extension of the $L_y = 1$ model, where world-lines are allowed to hop to diagonally opposite sites as illustrated in \cref{fig:conf2}. These diagonal bonds are given a tunable weight $J$. Thus, this extended model is characterized by two independent free parameters, $U$ and $J$.  \looseness-1

\begin{table}[!htb]
\centering
\renewcommand{\arraystretch}{1.4}
\setlength{\tabcolsep}{4pt}
\begin{tabular}{c|ccccc}
\TopRule
$(U,J)$ & $L_a$ & $\xi$($L_a$) & $\xi(L_a)/L_a$ &$\xi_\infty$ & $\xi\ (L_a)/\xi_\infty$ \\ 
\MidRule
(0.1,0) & 96 & 12.52(2) & 0.13 & 12.66(3) & 0.99 \\
(0,3.0) & 16 & 9.53(2) & 0.59 & 37.3(2) & 0.25\\
(0,5.0) & 12 & 7.97(1) & 0.66 & 61(1) & 0.13\\
\BotRule
\end{tabular}
\caption{\label{tab:correft} Correlation lengths achieved via tuning the couplings $U$ and $J$ in the single layer extended model. The various quantities shown are analogous to those in \cref{tab:corrDth}.}
\end{table}

The results for the step scaling function for this extended model are shown in \cref{fig:qeft}. The values for 
$L_a$, $\xi(L_a)$ and $\xi_\infty$ are given in \cref{tab:correft}, 
in analogy with those for the D-theory in \cref{tab:corrDth}. We find that this one-layer model can be tuned closer to the continuum limit by decreasing $U$ and increasing $J$. In particular when $U=0,J=5$, the correlation length $\xi_\infty$ reaches approximately $60$ in lattice units. 
Although we cannot approach closer and closer to the continuum limit, {i.e., \(\xi(L_a)/\xi_\infty = 0\),} we reached $\xi(L_a)/\xi_\infty \approx 0.13$ at $U=0,J=5$, which overlaps with the perturbative region. A further advantage of this two-dimensional model is that since $U=0$, the Hilbert space of the EFT is only four dimensional, and hence needs only two qubits per site to implement.

\section{Conclusion}
\label{sec:5}

In this work, we explored a space-time symmetric qubit regularization of the asymptotically free $O(4)$ QFT. We showed that we can reproduce the step-scaling function of the theory using a model with a five-dimensional Hilbert space per lattice site using the D-theory formalism by increasing the number of layers of two-dimensional lattices. With $L_y$ layers, this approach can be implemented using $3 L_y$ qubits per two-dimensional lattice site. \looseness-1

We then argued that it would be useful to view qubit regularization as constructing an effective field theory instead. This means that as long as a qubit regularized model can be tuned to reach close enough to the continuum limit for the step scaling function to make contact with perturbation theory, it can still be used to extract interesting physics. 
Noting that in the case of the $O(4)$ model, the two-loop perturbation theory matches the non-perturbative step scaling function very well in the regime $\xi(L)/L > 0.5$, we constructed a two coupling model in \cref{sec:4} and showed that with $U=0,J=5$ one obtains an excellent qubit-regularized model for the $O(4)$ field theory since it reaches $\xi(L_a)/L_a \approx 0.66$.  This model can be implemented with only two qubits per two-dimensional lattice site.

\section*{Acknowledgments}
HS would like to thank Stephan Caspar for useful conversations. The material presented here is based on work supported by the U.S. Department of Energy, Office of Science---High Energy Physics Contract KA2401032 (Triad National Security, LLC Contract Grant No. 89233218CNA000001) to Los Alamos National Laboratory. S.C. is supported by a Duke subcontract of this grant. S.C is also supported in part by the U.S. Department of Energy, Office of Science, Nuclear Physics program under Award No. DE-FG02-05ER41368.
HS is supported in part by the DOE QuantISED program through the theory  consortium ``Intersections of QIS and Theoretical Particle Physics'' at Fermilab with Fermilab Subcontract No. 666484,
in part by the Institute for Nuclear Theory with US Department of Energy Grant DE-FG02-00ER41132, and 
in part by the U.S. Department of Energy, Office of Science, Office of Nuclear Physics, Inqubator for Quantum Simulation (IQuS) under Award Number DOE (NP) Award DE-SC0020970.

\bibliographystyle{apsrev4-2}
\bibliography{refs}

\begin{thebibliography}{34}%
\makeatletter
\providecommand \@ifxundefined [1]{%
 \@ifx{#1\undefined}
}%
\providecommand \@ifnum [1]{%
 \ifnum #1\expandafter \@firstoftwo
 \else \expandafter \@secondoftwo
 \fi
}%
\providecommand \@ifx [1]{%
 \ifx #1\expandafter \@firstoftwo
 \else \expandafter \@secondoftwo
 \fi
}%
\providecommand \natexlab [1]{#1}%
\providecommand \enquote  [1]{``#1''}%
\providecommand \bibnamefont  [1]{#1}%
\providecommand \bibfnamefont [1]{#1}%
\providecommand \citenamefont [1]{#1}%
\providecommand \href@noop [0]{\@secondoftwo}%
\providecommand \href [0]{\begingroup \@sanitize@url \@href}%
\providecommand \@href[1]{\@@startlink{#1}\@@href}%
\providecommand \@@href[1]{\endgroup#1\@@endlink}%
\providecommand \@sanitize@url [0]{\catcode `\\12\catcode `\$12\catcode
  `\&12\catcode `\#12\catcode `\^12\catcode `\_12\catcode `\%12\relax}%
\providecommand \@@startlink[1]{}%
\providecommand \@@endlink[0]{}%
\providecommand \url  [0]{\begingroup\@sanitize@url \@url }%
\providecommand \@url [1]{\endgroup\@href {#1}{\urlprefix }}%
\providecommand \urlprefix  [0]{URL }%
\providecommand \Eprint [0]{\href }%
\providecommand \doibase [0]{https://doi.org/}%
\providecommand \selectlanguage [0]{\@gobble}%
\providecommand \bibinfo  [0]{\@secondoftwo}%
\providecommand \bibfield  [0]{\@secondoftwo}%
\providecommand \translation [1]{[#1]}%
\providecommand \BibitemOpen [0]{}%
\providecommand \bibitemStop [0]{}%
\providecommand \bibitemNoStop [0]{.\EOS\space}%
\providecommand \EOS [0]{\spacefactor3000\relax}%
\providecommand \BibitemShut  [1]{\csname bibitem#1\endcsname}%
\let\auto@bib@innerbib\@empty
\bibitem [{\citenamefont {Wilson}(1974)}]{Wilson:1974sk}%
  \BibitemOpen
  \bibfield  {author} {\bibinfo {author} {\bibfnamefont {K.~G.}\ \bibnamefont
  {Wilson}},\ }\href {https://doi.org/10.1103/PhysRevD.10.2445} {\bibfield
  {journal} {\bibinfo  {journal} {Phys. Rev. D}\ }\textbf {\bibinfo {volume}
  {10}},\ \bibinfo {pages} {2445} (\bibinfo {year} {1974})}\BibitemShut
  {NoStop}%
\bibitem [{\citenamefont {Kogut}(1979)}]{Kogut:1979wt}%
  \BibitemOpen
  \bibfield  {author} {\bibinfo {author} {\bibfnamefont {J.~B.}\ \bibnamefont
  {Kogut}},\ }\href {https://doi.org/10.1103/RevModPhys.51.659} {\bibfield
  {journal} {\bibinfo  {journal} {Rev. Mod. Phys.}\ }\textbf {\bibinfo {volume}
  {51}},\ \bibinfo {pages} {659} (\bibinfo {year} {1979})}\BibitemShut
  {NoStop}%
\bibitem [{\citenamefont {Jordan}\ \emph {et~al.}(2012)\citenamefont {Jordan},
  \citenamefont {Lee},\ and\ \citenamefont {Preskill}}]{Jordan:2011ne}%
  \BibitemOpen
  \bibfield  {author} {\bibinfo {author} {\bibfnamefont {S.~P.}\ \bibnamefont
  {Jordan}}, \bibinfo {author} {\bibfnamefont {K.~S.~M.}\ \bibnamefont {Lee}},\
  and\ \bibinfo {author} {\bibfnamefont {J.}~\bibnamefont {Preskill}},\ }\href
  {https://doi.org/10.1126/science.1217069} {\bibfield  {journal} {\bibinfo
  {journal} {Science}\ }\textbf {\bibinfo {volume} {336}},\ \bibinfo {pages}
  {1130} (\bibinfo {year} {2012})},\ \Eprint {https://arxiv.org/abs/1111.3633}
  {arXiv:1111.3633 [quant-ph]} \BibitemShut {NoStop}%
\bibitem [{\citenamefont {Casanova}\ \emph {et~al.}(2011)\citenamefont
  {Casanova}, \citenamefont {Lamata}, \citenamefont {Egusquiza}, \citenamefont
  {Gerritsma}, \citenamefont {Roos}, \citenamefont {Garcia-Ripoll},\ and\
  \citenamefont {Solano}}]{Casanova:2011wh}%
  \BibitemOpen
  \bibfield  {author} {\bibinfo {author} {\bibfnamefont {J.}~\bibnamefont
  {Casanova}}, \bibinfo {author} {\bibfnamefont {L.}~\bibnamefont {Lamata}},
  \bibinfo {author} {\bibfnamefont {I.~L.}\ \bibnamefont {Egusquiza}}, \bibinfo
  {author} {\bibfnamefont {R.}~\bibnamefont {Gerritsma}}, \bibinfo {author}
  {\bibfnamefont {C.~F.}\ \bibnamefont {Roos}}, \bibinfo {author}
  {\bibfnamefont {J.~J.}\ \bibnamefont {Garcia-Ripoll}},\ and\ \bibinfo
  {author} {\bibfnamefont {E.}~\bibnamefont {Solano}},\ }\href
  {https://doi.org/10.1103/PhysRevLett.107.260501} {\bibfield  {journal}
  {\bibinfo  {journal} {Phys. Rev. Lett.}\ }\textbf {\bibinfo {volume} {107}},\
  \bibinfo {pages} {260501} (\bibinfo {year} {2011})},\ \Eprint
  {https://arxiv.org/abs/1107.5233} {arXiv:1107.5233 [quant-ph]} \BibitemShut
  {NoStop}%
\bibitem [{\citenamefont {Casanova}\ \emph {et~al.}(2012)\citenamefont
  {Casanova}, \citenamefont {Mezzacapo}, \citenamefont {Lamata},\ and\
  \citenamefont {Solano}}]{Casanova:2012zz}%
  \BibitemOpen
  \bibfield  {author} {\bibinfo {author} {\bibfnamefont {J.}~\bibnamefont
  {Casanova}}, \bibinfo {author} {\bibfnamefont {A.}~\bibnamefont {Mezzacapo}},
  \bibinfo {author} {\bibfnamefont {L.}~\bibnamefont {Lamata}},\ and\ \bibinfo
  {author} {\bibfnamefont {E.}~\bibnamefont {Solano}},\ }\href
  {https://doi.org/10.1103/PhysRevLett.108.190502} {\bibfield  {journal}
  {\bibinfo  {journal} {Phys. Rev. Lett.}\ }\textbf {\bibinfo {volume} {108}},\
  \bibinfo {pages} {190502} (\bibinfo {year} {2012})},\ \Eprint
  {https://arxiv.org/abs/1110.3730} {arXiv:1110.3730 [quant-ph]} \BibitemShut
  {NoStop}%
\bibitem [{\citenamefont {O'Malley}\ \emph {et~al.}(2016)\citenamefont
  {O'Malley}, \citenamefont {Babbush}, \citenamefont {Kivlichan}, \citenamefont
  {Romero}, \citenamefont {McClean}, \citenamefont {Barends}, \citenamefont
  {Kelly}, \citenamefont {Roushan}, \citenamefont {Tranter}, \citenamefont
  {Ding}, \citenamefont {Campbell}, \citenamefont {Chen}, \citenamefont {Chen},
  \citenamefont {Chiaro}, \citenamefont {Dunsworth}, \citenamefont {Fowler},
  \citenamefont {Jeffrey}, \citenamefont {Lucero}, \citenamefont {Megrant},
  \citenamefont {Mutus}, \citenamefont {Neeley}, \citenamefont {Neill},
  \citenamefont {Quintana}, \citenamefont {Sank}, \citenamefont {Vainsencher},
  \citenamefont {Wenner}, \citenamefont {White}, \citenamefont {Coveney},
  \citenamefont {Love}, \citenamefont {Neven}, \citenamefont {Aspuru-Guzik},\
  and\ \citenamefont {Martinis}}]{PhysRevX.6.031007}%
  \BibitemOpen
  \bibfield  {author} {\bibinfo {author} {\bibfnamefont {P.~J.~J.}\
  \bibnamefont {O'Malley}}, \bibinfo {author} {\bibfnamefont {R.}~\bibnamefont
  {Babbush}}, \bibinfo {author} {\bibfnamefont {I.~D.}\ \bibnamefont
  {Kivlichan}}, \bibinfo {author} {\bibfnamefont {J.}~\bibnamefont {Romero}},
  \bibinfo {author} {\bibfnamefont {J.~R.}\ \bibnamefont {McClean}}, \bibinfo
  {author} {\bibfnamefont {R.}~\bibnamefont {Barends}}, \bibinfo {author}
  {\bibfnamefont {J.}~\bibnamefont {Kelly}}, \bibinfo {author} {\bibfnamefont
  {P.}~\bibnamefont {Roushan}}, \bibinfo {author} {\bibfnamefont
  {A.}~\bibnamefont {Tranter}}, \bibinfo {author} {\bibfnamefont
  {N.}~\bibnamefont {Ding}}, \bibinfo {author} {\bibfnamefont {B.}~\bibnamefont
  {Campbell}}, \bibinfo {author} {\bibfnamefont {Y.}~\bibnamefont {Chen}},
  \bibinfo {author} {\bibfnamefont {Z.}~\bibnamefont {Chen}}, \bibinfo {author}
  {\bibfnamefont {B.}~\bibnamefont {Chiaro}}, \bibinfo {author} {\bibfnamefont
  {A.}~\bibnamefont {Dunsworth}}, \bibinfo {author} {\bibfnamefont {A.~G.}\
  \bibnamefont {Fowler}}, \bibinfo {author} {\bibfnamefont {E.}~\bibnamefont
  {Jeffrey}}, \bibinfo {author} {\bibfnamefont {E.}~\bibnamefont {Lucero}},
  \bibinfo {author} {\bibfnamefont {A.}~\bibnamefont {Megrant}}, \bibinfo
  {author} {\bibfnamefont {J.~Y.}\ \bibnamefont {Mutus}}, \bibinfo {author}
  {\bibfnamefont {M.}~\bibnamefont {Neeley}}, \bibinfo {author} {\bibfnamefont
  {C.}~\bibnamefont {Neill}}, \bibinfo {author} {\bibfnamefont
  {C.}~\bibnamefont {Quintana}}, \bibinfo {author} {\bibfnamefont
  {D.}~\bibnamefont {Sank}}, \bibinfo {author} {\bibfnamefont {A.}~\bibnamefont
  {Vainsencher}}, \bibinfo {author} {\bibfnamefont {J.}~\bibnamefont {Wenner}},
  \bibinfo {author} {\bibfnamefont {T.~C.}\ \bibnamefont {White}}, \bibinfo
  {author} {\bibfnamefont {P.~V.}\ \bibnamefont {Coveney}}, \bibinfo {author}
  {\bibfnamefont {P.~J.}\ \bibnamefont {Love}}, \bibinfo {author}
  {\bibfnamefont {H.}~\bibnamefont {Neven}}, \bibinfo {author} {\bibfnamefont
  {A.}~\bibnamefont {Aspuru-Guzik}},\ and\ \bibinfo {author} {\bibfnamefont
  {J.~M.}\ \bibnamefont {Martinis}},\ }\href
  {https://doi.org/10.1103/PhysRevX.6.031007} {\bibfield  {journal} {\bibinfo
  {journal} {Phys. Rev. X}\ }\textbf {\bibinfo {volume} {6}},\ \bibinfo {pages}
  {031007} (\bibinfo {year} {2016})}\BibitemShut {NoStop}%
\bibitem [{\citenamefont {Klco}\ \emph {et~al.}(2018)\citenamefont {Klco},
  \citenamefont {Dumitrescu}, \citenamefont {McCaskey}, \citenamefont {Morris},
  \citenamefont {Pooser}, \citenamefont {Sanz}, \citenamefont {Solano},
  \citenamefont {Lougovski},\ and\ \citenamefont {Savage}}]{Klco:2018kyo}%
  \BibitemOpen
  \bibfield  {author} {\bibinfo {author} {\bibfnamefont {N.}~\bibnamefont
  {Klco}}, \bibinfo {author} {\bibfnamefont {E.~F.}\ \bibnamefont
  {Dumitrescu}}, \bibinfo {author} {\bibfnamefont {A.~J.}\ \bibnamefont
  {McCaskey}}, \bibinfo {author} {\bibfnamefont {T.~D.}\ \bibnamefont
  {Morris}}, \bibinfo {author} {\bibfnamefont {R.~C.}\ \bibnamefont {Pooser}},
  \bibinfo {author} {\bibfnamefont {M.}~\bibnamefont {Sanz}}, \bibinfo {author}
  {\bibfnamefont {E.}~\bibnamefont {Solano}}, \bibinfo {author} {\bibfnamefont
  {P.}~\bibnamefont {Lougovski}},\ and\ \bibinfo {author} {\bibfnamefont
  {M.~J.}\ \bibnamefont {Savage}},\ }\href
  {https://doi.org/10.1103/PhysRevA.98.032331} {\bibfield  {journal} {\bibinfo
  {journal} {Phys. Rev.}\ }\textbf {\bibinfo {volume} {A98}},\ \bibinfo {pages}
  {032331} (\bibinfo {year} {2018})},\ \Eprint
  {https://arxiv.org/abs/1803.03326} {arXiv:1803.03326 [quant-ph]} \BibitemShut
  {NoStop}%
\bibitem [{\citenamefont {Klco}\ \emph {et~al.}(2020)\citenamefont {Klco},
  \citenamefont {Stryker},\ and\ \citenamefont {Savage}}]{Klco:2019evd}%
  \BibitemOpen
  \bibfield  {author} {\bibinfo {author} {\bibfnamefont {N.}~\bibnamefont
  {Klco}}, \bibinfo {author} {\bibfnamefont {J.~R.}\ \bibnamefont {Stryker}},\
  and\ \bibinfo {author} {\bibfnamefont {M.~J.}\ \bibnamefont {Savage}},\
  }\href {https://doi.org/10.1103/PhysRevD.101.074512} {\bibfield  {journal}
  {\bibinfo  {journal} {Phys. Rev. D}\ }\textbf {\bibinfo {volume} {101}},\
  \bibinfo {pages} {074512} (\bibinfo {year} {2020})},\ \Eprint
  {https://arxiv.org/abs/1908.06935} {arXiv:1908.06935 [quant-ph]} \BibitemShut
  {NoStop}%
\bibitem [{\citenamefont {Alexandru}\ \emph
  {et~al.}(2019{\natexlab{a}})\citenamefont {Alexandru}, \citenamefont
  {Bedaque}, \citenamefont {Harmalkar}, \citenamefont {Lamm}, \citenamefont
  {Lawrence},\ and\ \citenamefont {Warrington}}]{Alexandru:2019nsa}%
  \BibitemOpen
  \bibfield  {author} {\bibinfo {author} {\bibfnamefont {A.}~\bibnamefont
  {Alexandru}}, \bibinfo {author} {\bibfnamefont {P.~F.}\ \bibnamefont
  {Bedaque}}, \bibinfo {author} {\bibfnamefont {S.}~\bibnamefont {Harmalkar}},
  \bibinfo {author} {\bibfnamefont {H.}~\bibnamefont {Lamm}}, \bibinfo {author}
  {\bibfnamefont {S.}~\bibnamefont {Lawrence}},\ and\ \bibinfo {author}
  {\bibfnamefont {N.~C.}\ \bibnamefont {Warrington}} (\bibinfo {collaboration}
  {NuQS}),\ }\href {https://doi.org/10.1103/PhysRevD.100.114501} {\bibfield
  {journal} {\bibinfo  {journal} {Phys. Rev. D}\ }\textbf {\bibinfo {volume}
  {100}},\ \bibinfo {pages} {114501} (\bibinfo {year} {2019}{\natexlab{a}})},\
  \Eprint {https://arxiv.org/abs/1906.11213} {arXiv:1906.11213 [hep-lat]}
  \BibitemShut {NoStop}%
\bibitem [{\citenamefont {Ba\~nuls}\ \emph {et~al.}(2020)\citenamefont
  {Ba\~nuls} \emph {et~al.}}]{Banuls:2019bmf}%
  \BibitemOpen
  \bibfield  {author} {\bibinfo {author} {\bibfnamefont {M.}~\bibnamefont
  {Ba\~nuls}} \emph {et~al.},\ }\href
  {https://doi.org/10.1140/epjd/e2020-100571-8} {\bibfield  {journal} {\bibinfo
   {journal} {Eur. Phys. J. D}\ }\textbf {\bibinfo {volume} {74}},\ \bibinfo
  {pages} {165} (\bibinfo {year} {2020})},\ \Eprint
  {https://arxiv.org/abs/1911.00003} {arXiv:1911.00003 [quant-ph]} \BibitemShut
  {NoStop}%
\bibitem [{\citenamefont {Ba\~nuls}\ and\ \citenamefont
  {Cichy}(2020)}]{Banuls:2019rao}%
  \BibitemOpen
  \bibfield  {author} {\bibinfo {author} {\bibfnamefont {M.~C.}\ \bibnamefont
  {Ba\~nuls}}\ and\ \bibinfo {author} {\bibfnamefont {K.}~\bibnamefont
  {Cichy}},\ }\href {https://doi.org/10.1088/1361-6633/ab6311} {\bibfield
  {journal} {\bibinfo  {journal} {Rept. Prog. Phys.}\ }\textbf {\bibinfo
  {volume} {83}},\ \bibinfo {pages} {024401} (\bibinfo {year} {2020})},\
  \Eprint {https://arxiv.org/abs/1910.00257} {arXiv:1910.00257 [hep-lat]}
  \BibitemShut {NoStop}%
\bibitem [{\citenamefont {Alexandru}\ \emph
  {et~al.}(2019{\natexlab{b}})\citenamefont {Alexandru}, \citenamefont
  {Bedaque}, \citenamefont {Lamm},\ and\ \citenamefont
  {Lawrence}}]{Alexandru:2019ozf}%
  \BibitemOpen
  \bibfield  {author} {\bibinfo {author} {\bibfnamefont {A.}~\bibnamefont
  {Alexandru}}, \bibinfo {author} {\bibfnamefont {P.~F.}\ \bibnamefont
  {Bedaque}}, \bibinfo {author} {\bibfnamefont {H.}~\bibnamefont {Lamm}},\ and\
  \bibinfo {author} {\bibfnamefont {S.}~\bibnamefont {Lawrence}} (\bibinfo
  {collaboration} {NuQS}),\ }\href
  {https://doi.org/10.1103/PhysRevLett.123.090501} {\bibfield  {journal}
  {\bibinfo  {journal} {Phys. Rev. Lett.}\ }\textbf {\bibinfo {volume} {123}},\
  \bibinfo {pages} {090501} (\bibinfo {year} {2019}{\natexlab{b}})},\ \Eprint
  {https://arxiv.org/abs/1903.06577} {arXiv:1903.06577 [hep-lat]} \BibitemShut
  {NoStop}%
\bibitem [{\citenamefont {Raychowdhury}\ and\ \citenamefont
  {Stryker}(2020)}]{Raychowdhury:2019iki}%
  \BibitemOpen
  \bibfield  {author} {\bibinfo {author} {\bibfnamefont {I.}~\bibnamefont
  {Raychowdhury}}\ and\ \bibinfo {author} {\bibfnamefont {J.~R.}\ \bibnamefont
  {Stryker}},\ }\href {https://doi.org/10.1103/PhysRevD.101.114502} {\bibfield
  {journal} {\bibinfo  {journal} {Phys. Rev. D}\ }\textbf {\bibinfo {volume}
  {101}},\ \bibinfo {pages} {114502} (\bibinfo {year} {2020})},\ \Eprint
  {https://arxiv.org/abs/1912.06133} {arXiv:1912.06133 [hep-lat]} \BibitemShut
  {NoStop}%
\bibitem [{\citenamefont {Bender}\ \emph {et~al.}(2020)\citenamefont {Bender},
  \citenamefont {Emonts}, \citenamefont {Zohar},\ and\ \citenamefont
  {Cirac}}]{Bender:2020jgr}%
  \BibitemOpen
  \bibfield  {author} {\bibinfo {author} {\bibfnamefont {J.}~\bibnamefont
  {Bender}}, \bibinfo {author} {\bibfnamefont {P.}~\bibnamefont {Emonts}},
  \bibinfo {author} {\bibfnamefont {E.}~\bibnamefont {Zohar}},\ and\ \bibinfo
  {author} {\bibfnamefont {J.~I.}\ \bibnamefont {Cirac}},\ }\href
  {https://doi.org/10.1103/PhysRevResearch.2.043145} {\bibfield  {journal}
  {\bibinfo  {journal} {Phys. Rev. Res.}\ }\textbf {\bibinfo {volume} {2}},\
  \bibinfo {pages} {043145} (\bibinfo {year} {2020})},\ \Eprint
  {https://arxiv.org/abs/2006.10038} {arXiv:2006.10038 [hep-th]} \BibitemShut
  {NoStop}%
\bibitem [{\citenamefont {Alexandru}\ \emph {et~al.}(2021)\citenamefont
  {Alexandru}, \citenamefont {Bedaque}, \citenamefont {Carosso},\ and\
  \citenamefont {Sheng}}]{Alexandru:2021xkf}%
  \BibitemOpen
  \bibfield  {author} {\bibinfo {author} {\bibfnamefont {A.}~\bibnamefont
  {Alexandru}}, \bibinfo {author} {\bibfnamefont {P.~F.}\ \bibnamefont
  {Bedaque}}, \bibinfo {author} {\bibfnamefont {A.}~\bibnamefont {Carosso}},\
  and\ \bibinfo {author} {\bibfnamefont {A.}~\bibnamefont {Sheng}},\
  }\href@noop {} {\bibinfo {title} {{Universality of a truncated sigma-model}}}
  (\bibinfo {year} {2021}),\ \Eprint {https://arxiv.org/abs/2109.07500}
  {arXiv:2109.07500 [hep-lat]} \BibitemShut {NoStop}%
\bibitem [{\citenamefont {Singh}\ and\ \citenamefont
  {Chandrasekharan}(2019)}]{Singh:2019uwd}%
  \BibitemOpen
  \bibfield  {author} {\bibinfo {author} {\bibfnamefont {H.}~\bibnamefont
  {Singh}}\ and\ \bibinfo {author} {\bibfnamefont {S.}~\bibnamefont
  {Chandrasekharan}},\ }\href {https://doi.org/10.1103/PhysRevD.100.054505}
  {\bibfield  {journal} {\bibinfo  {journal} {Phys. Rev.}\ }\textbf {\bibinfo
  {volume} {D100}},\ \bibinfo {pages} {054505} (\bibinfo {year} {2019})},\
  \Eprint {https://arxiv.org/abs/1905.13204} {arXiv:1905.13204 [hep-lat]}
  \BibitemShut {NoStop}%
\bibitem [{\citenamefont {Zhang}\ \emph {et~al.}(2021)\citenamefont {Zhang},
  \citenamefont {Meurice},\ and\ \citenamefont {Tsai}}]{PhysRevB.103.245137}%
  \BibitemOpen
  \bibfield  {author} {\bibinfo {author} {\bibfnamefont {J.}~\bibnamefont
  {Zhang}}, \bibinfo {author} {\bibfnamefont {Y.}~\bibnamefont {Meurice}},\
  and\ \bibinfo {author} {\bibfnamefont {S.-W.}\ \bibnamefont {Tsai}},\ }\href
  {https://doi.org/10.1103/PhysRevB.103.245137} {\bibfield  {journal} {\bibinfo
   {journal} {Phys. Rev. B}\ }\textbf {\bibinfo {volume} {103}},\ \bibinfo
  {pages} {245137} (\bibinfo {year} {2021})}\BibitemShut {NoStop}%
\bibitem [{\citenamefont {Singh}(2019)}]{Singh:2019jog}%
  \BibitemOpen
  \bibfield  {author} {\bibinfo {author} {\bibfnamefont {H.}~\bibnamefont
  {Singh}},\ }\href@noop {} {\bibinfo {title} {{Qubit $O(N)$ nonlinear sigma
  models}}} (\bibinfo {year} {2019}),\ \Eprint
  {https://arxiv.org/abs/1911.12353} {arXiv:1911.12353 [hep-lat]} \BibitemShut
  {NoStop}%
\bibitem [{\citenamefont {Sandvik}\ and\ \citenamefont
  {Scalapino}(1994)}]{PhysRevLett.72.2777}%
  \BibitemOpen
  \bibfield  {author} {\bibinfo {author} {\bibfnamefont {A.~W.}\ \bibnamefont
  {Sandvik}}\ and\ \bibinfo {author} {\bibfnamefont {D.~J.}\ \bibnamefont
  {Scalapino}},\ }\href {https://doi.org/10.1103/PhysRevLett.72.2777}
  {\bibfield  {journal} {\bibinfo  {journal} {Phys. Rev. Lett.}\ }\textbf
  {\bibinfo {volume} {72}},\ \bibinfo {pages} {2777} (\bibinfo {year}
  {1994})}\BibitemShut {NoStop}%
\bibitem [{\citenamefont {Brower}\ \emph {et~al.}(2004)\citenamefont {Brower},
  \citenamefont {Chandrasekharan}, \citenamefont {Riederer},\ and\
  \citenamefont {Wiese}}]{Brower:2003vy}%
  \BibitemOpen
  \bibfield  {author} {\bibinfo {author} {\bibfnamefont {R.}~\bibnamefont
  {Brower}}, \bibinfo {author} {\bibfnamefont {S.}~\bibnamefont
  {Chandrasekharan}}, \bibinfo {author} {\bibfnamefont {S.}~\bibnamefont
  {Riederer}},\ and\ \bibinfo {author} {\bibfnamefont {U.}~\bibnamefont
  {Wiese}},\ }\href {https://doi.org/10.1016/j.nuclphysb.2004.06.007}
  {\bibfield  {journal} {\bibinfo  {journal} {Nucl. Phys. B}\ }\textbf
  {\bibinfo {volume} {693}},\ \bibinfo {pages} {149} (\bibinfo {year}
  {2004})},\ \Eprint {https://arxiv.org/abs/hep-lat/0309182}
  {arXiv:hep-lat/0309182} \BibitemShut {NoStop}%
\bibitem [{\citenamefont {Wiese}(2006)}]{Wiese:2006kp}%
  \BibitemOpen
  \bibfield  {author} {\bibinfo {author} {\bibfnamefont {U.~J.}\ \bibnamefont
  {Wiese}},\ }\bibfield  {booktitle} {\emph {\bibinfo {booktitle} {{Hadron
  physics, proceedings of the Workshop on Computational Hadron Physics,
  University of Cyprus, Nicosia, Cyprus, 14-17 September 2005}}},\ }\href
  {https://doi.org/10.1016/j.nuclphysbps.2006.01.027} {\bibfield  {journal}
  {\bibinfo  {journal} {Nucl. Phys. Proc. Suppl.}\ }\textbf {\bibinfo {volume}
  {153}},\ \bibinfo {pages} {336} (\bibinfo {year} {2006})}\BibitemShut
  {NoStop}%
\bibitem [{\citenamefont {Evans}\ \emph {et~al.}(2018)\citenamefont {Evans},
  \citenamefont {Gerber}, \citenamefont {Hornung},\ and\ \citenamefont
  {Wiese}}]{Evans:2018njs}%
  \BibitemOpen
  \bibfield  {author} {\bibinfo {author} {\bibfnamefont {W.}~\bibnamefont
  {Evans}}, \bibinfo {author} {\bibfnamefont {U.}~\bibnamefont {Gerber}},
  \bibinfo {author} {\bibfnamefont {M.}~\bibnamefont {Hornung}},\ and\ \bibinfo
  {author} {\bibfnamefont {U.-J.}\ \bibnamefont {Wiese}},\ }\href
  {https://doi.org/10.1016/j.aop.2018.09.002} {\bibfield  {journal} {\bibinfo
  {journal} {Annals Phys.}\ }\textbf {\bibinfo {volume} {398}},\ \bibinfo
  {pages} {94} (\bibinfo {year} {2018})},\ \Eprint
  {https://arxiv.org/abs/1803.04767} {arXiv:1803.04767 [hep-lat]} \BibitemShut
  {NoStop}%
\bibitem [{\citenamefont {Bietenholz}\ \emph {et~al.}(2003)\citenamefont
  {Bietenholz}, \citenamefont {Gfeller},\ and\ \citenamefont
  {Wiese}}]{Bietenholz:2003wa}%
  \BibitemOpen
  \bibfield  {author} {\bibinfo {author} {\bibfnamefont {W.}~\bibnamefont
  {Bietenholz}}, \bibinfo {author} {\bibfnamefont {A.}~\bibnamefont
  {Gfeller}},\ and\ \bibinfo {author} {\bibfnamefont {U.~J.}\ \bibnamefont
  {Wiese}},\ }\href {https://doi.org/10.1088/1126-6708/2003/10/018} {\bibfield
  {journal} {\bibinfo  {journal} {JHEP}\ }\textbf {\bibinfo {volume}
  {10}}\bibfield  {number} {\bibinfo  {number} { (2003)},\ \bibinfo {pages}
  {018}},\ }\Eprint {https://arxiv.org/abs/hep-th/0309162}
  {arXiv:hep-th/0309162} \BibitemShut {NoStop}%
\bibitem [{\citenamefont {Beard}\ \emph {et~al.}(2005)\citenamefont {Beard},
  \citenamefont {Pepe}, \citenamefont {Riederer},\ and\ \citenamefont
  {Wiese}}]{Beard:2004jr}%
  \BibitemOpen
  \bibfield  {author} {\bibinfo {author} {\bibfnamefont {B.}~\bibnamefont
  {Beard}}, \bibinfo {author} {\bibfnamefont {M.}~\bibnamefont {Pepe}},
  \bibinfo {author} {\bibfnamefont {S.}~\bibnamefont {Riederer}},\ and\
  \bibinfo {author} {\bibfnamefont {U.}~\bibnamefont {Wiese}},\ }\href
  {https://doi.org/10.1103/PhysRevLett.94.010603} {\bibfield  {journal}
  {\bibinfo  {journal} {Phys. Rev. Lett.}\ }\textbf {\bibinfo {volume} {94}},\
  \bibinfo {pages} {010603} (\bibinfo {year} {2005})},\ \Eprint
  {https://arxiv.org/abs/hep-lat/0406040} {arXiv:hep-lat/0406040} \BibitemShut
  {NoStop}%
\bibitem [{\citenamefont {Luscher}\ \emph {et~al.}(1991)\citenamefont
  {Luscher}, \citenamefont {Weisz},\ and\ \citenamefont
  {Wolff}}]{Luscher:1991wu}%
  \BibitemOpen
  \bibfield  {author} {\bibinfo {author} {\bibfnamefont {M.}~\bibnamefont
  {Luscher}}, \bibinfo {author} {\bibfnamefont {P.}~\bibnamefont {Weisz}},\
  and\ \bibinfo {author} {\bibfnamefont {U.}~\bibnamefont {Wolff}},\ }\href
  {https://doi.org/10.1016/0550-3213(91)90298-C} {\bibfield  {journal}
  {\bibinfo  {journal} {Nucl. Phys.}\ }\textbf {\bibinfo {volume} {B359}},\
  \bibinfo {pages} {221} (\bibinfo {year} {1991})}\BibitemShut {NoStop}%
\bibitem [{\citenamefont {Cooper}\ \emph {et~al.}(1982)\citenamefont {Cooper},
  \citenamefont {Freedman},\ and\ \citenamefont {Preston}}]{Cooper:1982nn}%
  \BibitemOpen
  \bibfield  {author} {\bibinfo {author} {\bibfnamefont {F.}~\bibnamefont
  {Cooper}}, \bibinfo {author} {\bibfnamefont {B.}~\bibnamefont {Freedman}},\
  and\ \bibinfo {author} {\bibfnamefont {D.}~\bibnamefont {Preston}},\ }\href
  {https://doi.org/10.1016/0550-3213(82)90240-1} {\bibfield  {journal}
  {\bibinfo  {journal} {Nucl. Phys. B}\ }\textbf {\bibinfo {volume} {210}},\
  \bibinfo {pages} {210} (\bibinfo {year} {1982})}\BibitemShut {NoStop}%
\bibitem [{\citenamefont {Gupta}\ and\ \citenamefont
  {Baillie}(1992)}]{PhysRevB.45.2883}%
  \BibitemOpen
  \bibfield  {author} {\bibinfo {author} {\bibfnamefont {R.}~\bibnamefont
  {Gupta}}\ and\ \bibinfo {author} {\bibfnamefont {C.~F.}\ \bibnamefont
  {Baillie}},\ }\href {https://doi.org/10.1103/PhysRevB.45.2883} {\bibfield
  {journal} {\bibinfo  {journal} {Phys. Rev. B}\ }\textbf {\bibinfo {volume}
  {45}},\ \bibinfo {pages} {2883} (\bibinfo {year} {1992})}\BibitemShut
  {NoStop}%
\bibitem [{\citenamefont {Caracciolo}\ \emph {et~al.}(1993)\citenamefont
  {Caracciolo}, \citenamefont {Edwards}, \citenamefont {Pelissetto},\ and\
  \citenamefont {Sokal}}]{Caracciolo:1992nh}%
  \BibitemOpen
  \bibfield  {author} {\bibinfo {author} {\bibfnamefont {S.}~\bibnamefont
  {Caracciolo}}, \bibinfo {author} {\bibfnamefont {R.~G.}\ \bibnamefont
  {Edwards}}, \bibinfo {author} {\bibfnamefont {A.}~\bibnamefont
  {Pelissetto}},\ and\ \bibinfo {author} {\bibfnamefont {A.~D.}\ \bibnamefont
  {Sokal}},\ }\href {https://doi.org/10.1016/0550-3213(93)90044-P} {\bibfield
  {journal} {\bibinfo  {journal} {Nucl. Phys. B}\ }\textbf {\bibinfo {volume}
  {403}},\ \bibinfo {pages} {475} (\bibinfo {year} {1993})},\ \Eprint
  {https://arxiv.org/abs/hep-lat/9205005} {arXiv:hep-lat/9205005} \BibitemShut
  {NoStop}%
\bibitem [{\citenamefont {Wolff}(1989)}]{Wolff:1988uh}%
  \BibitemOpen
  \bibfield  {author} {\bibinfo {author} {\bibfnamefont {U.}~\bibnamefont
  {Wolff}},\ }\href {https://doi.org/10.1103/PhysRevLett.62.361} {\bibfield
  {journal} {\bibinfo  {journal} {Phys. Rev. Lett.}\ }\textbf {\bibinfo
  {volume} {62}},\ \bibinfo {pages} {361} (\bibinfo {year} {1989})}\BibitemShut
  {NoStop}%
\bibitem [{\citenamefont {Wolff}(1990)}]{Wolff:1989hv}%
  \BibitemOpen
  \bibfield  {author} {\bibinfo {author} {\bibfnamefont {U.}~\bibnamefont
  {Wolff}},\ }\href {https://doi.org/10.1016/0550-3213(90)90313-3} {\bibfield
  {journal} {\bibinfo  {journal} {Nucl. Phys. B}\ }\textbf {\bibinfo {volume}
  {334}},\ \bibinfo {pages} {581} (\bibinfo {year} {1990})}\BibitemShut
  {NoStop}%
\bibitem [{\citenamefont {Cecile}\ and\ \citenamefont
  {Chandrasekharan}(2008)}]{Cecile:2007dv}%
  \BibitemOpen
  \bibfield  {author} {\bibinfo {author} {\bibfnamefont {D.~J.}\ \bibnamefont
  {Cecile}}\ and\ \bibinfo {author} {\bibfnamefont {S.}~\bibnamefont
  {Chandrasekharan}},\ }\href {https://doi.org/10.1103/PhysRevD.77.014506}
  {\bibfield  {journal} {\bibinfo  {journal} {Phys. Rev.}\ }\textbf {\bibinfo
  {volume} {D77}},\ \bibinfo {pages} {014506} (\bibinfo {year} {2008})},\
  \Eprint {https://arxiv.org/abs/0708.0558} {arXiv:0708.0558 [hep-lat]}
  \BibitemShut {NoStop}%
\bibitem [{\citenamefont {Banerjee}\ \emph {et~al.}(2019)\citenamefont
  {Banerjee}, \citenamefont {Chandrasekharan}, \citenamefont {Orlando},\ and\
  \citenamefont {Reffert}}]{Banerjee:2019jpw}%
  \BibitemOpen
  \bibfield  {author} {\bibinfo {author} {\bibfnamefont {D.}~\bibnamefont
  {Banerjee}}, \bibinfo {author} {\bibfnamefont {S.}~\bibnamefont
  {Chandrasekharan}}, \bibinfo {author} {\bibfnamefont {D.}~\bibnamefont
  {Orlando}},\ and\ \bibinfo {author} {\bibfnamefont {S.}~\bibnamefont
  {Reffert}},\ }\href {https://doi.org/10.1103/PhysRevLett.123.051603}
  {\bibfield  {journal} {\bibinfo  {journal} {Phys. Rev. Lett.}\ }\textbf
  {\bibinfo {volume} {123}},\ \bibinfo {pages} {051603} (\bibinfo {year}
  {2019})},\ \Eprint {https://arxiv.org/abs/1902.09542} {arXiv:1902.09542
  [hep-lat]} \BibitemShut {NoStop}%
\bibitem [{\citenamefont {Caracciolo}\ \emph {et~al.}(1995)\citenamefont
  {Caracciolo}, \citenamefont {Edwards}, \citenamefont {Pelissetto},\ and\
  \citenamefont {Sokal}}]{Caracciolo:1994ud}%
  \BibitemOpen
  \bibfield  {author} {\bibinfo {author} {\bibfnamefont {S.}~\bibnamefont
  {Caracciolo}}, \bibinfo {author} {\bibfnamefont {R.~G.}\ \bibnamefont
  {Edwards}}, \bibinfo {author} {\bibfnamefont {A.}~\bibnamefont
  {Pelissetto}},\ and\ \bibinfo {author} {\bibfnamefont {A.~D.}\ \bibnamefont
  {Sokal}},\ }\href {https://doi.org/10.1103/PhysRevLett.75.1891} {\bibfield
  {journal} {\bibinfo  {journal} {Phys. Rev. Lett.}\ }\textbf {\bibinfo
  {volume} {75}},\ \bibinfo {pages} {1891} (\bibinfo {year} {1995})},\ \Eprint
  {https://arxiv.org/abs/hep-lat/9411009} {arXiv:hep-lat/9411009} \BibitemShut
  {NoStop}%
\bibitem [{\citenamefont {Bhattacharya}\ \emph {et~al.}(2021)\citenamefont
  {Bhattacharya}, \citenamefont {Buser}, \citenamefont {Chandrasekharan},
  \citenamefont {Gupta},\ and\ \citenamefont {Singh}}]{Bhattacharya:2020gpm}%
  \BibitemOpen
  \bibfield  {author} {\bibinfo {author} {\bibfnamefont {T.}~\bibnamefont
  {Bhattacharya}}, \bibinfo {author} {\bibfnamefont {A.~J.}\ \bibnamefont
  {Buser}}, \bibinfo {author} {\bibfnamefont {S.}~\bibnamefont
  {Chandrasekharan}}, \bibinfo {author} {\bibfnamefont {R.}~\bibnamefont
  {Gupta}},\ and\ \bibinfo {author} {\bibfnamefont {H.}~\bibnamefont {Singh}},\
  }\href {https://doi.org/10.1103/physrevlett.126.172001} {\bibfield  {journal}
  {\bibinfo  {journal} {Physical Review Letters}\ }\textbf {\bibinfo {volume}
  {126}},\ \bibinfo {pages} {172001} (\bibinfo {year} {2021})}\BibitemShut
  {NoStop}%
\end{thebibliography}%

\appendix
\section{Exact calculations on small lattices}

\begin{figure}[htbp]
\includegraphics[width=0.95\hsize]{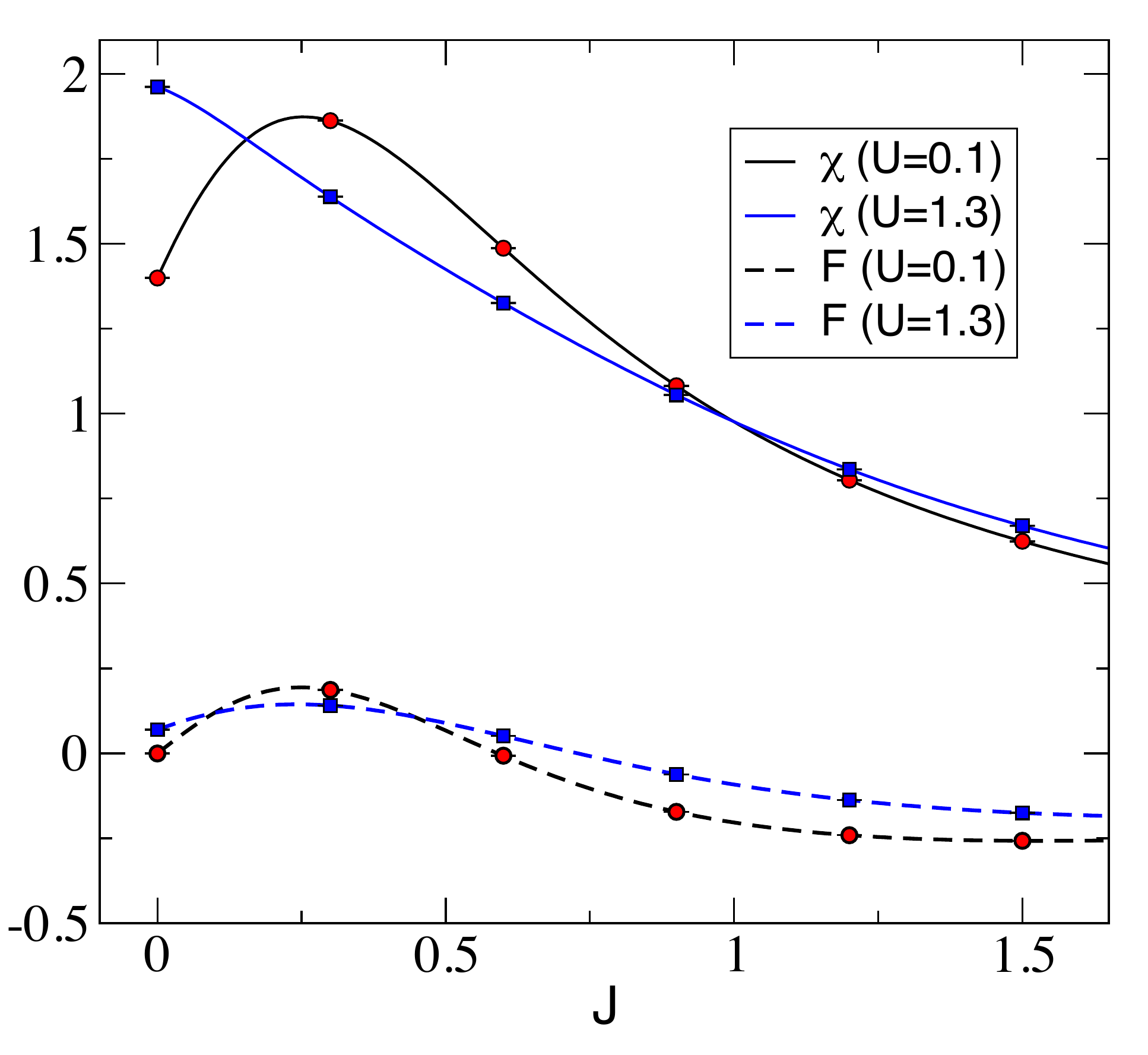}
\caption{Comparison of exact results with Monte Carlo results for the observables $\chi$ (solid lines) and $F$ (dashed lines) at $U=0.1$ (circles) and $U=1.3$ (squares) as a function of the coupling $J$.
\label{fig:exact}}
\end{figure}

The exact results for the extended model,  whose configurations are described in \cref{fig:conf2}, used to check the Monte Carlo implementation, are presented in this Appendix. On a $2\times 2$ lattice with periodic boundary conditions, the partition function of the model and the two observables $\chi$ and $F$ defined in \cref{eq:corrlen},  are given by
\begin{align}
Z &= \ U^4 + 16U^2 + 48J^2U^2 + 256UJ 
\nonumber \\
& \qquad \qquad \qquad \qquad + 576J^4 + 512J^2 + 96 \\
\chi &= 4 U^3 + 16 U^2 + 16 U^2J + 128 U J + 64 U 
\nonumber \\
& \qquad + 96 U J^2 + 384 J^3 + 256 J^2 + 384 J + 128 \\
F &=  4 U^3 - 16 U^2J + 96 U J^2 - 384 J^3 + 128 J.
\end{align}
In \cref{fig:exact} we plot the values of $\chi$ and $F$ obtained using the above relations as a function of $J$ at $U=0.1$ and $U=1.3$. For comparison, results obtained from our Monte Carlo algorithm are given at a few values of $J$.

\section{Monte Carlo Results}

In this appendix we give results for our observables $\chi$, $F$ obtained from our Monte Carlo calculations at various values of $L$. We also give results for $\chi(L)$ obtained from $\chi$ and $F$ using the relation \cref{eq:corrlen}. 

\begin{table*}[!htbp]
\renewcommand{\arraystretch}{1.4}
\setlength{\tabcolsep}{4pt}
\begin{tabular}{c|ccc|ccc|ccc}
\TopRule
 & \multicolumn{3}{c|}{$L_y=1$} &
\multicolumn{3}{c|}{$L_y=3$} & \multicolumn{3}{c}{$L_y=5$} \\ 
\TopRule
$L$ & $\chi$ & $F$ & $\xi(L)$ & $\chi$ & $F$ & $\xi(L)$ & $\chi$ & $F$ & $\xi(L)$ \\
\MidRule
4 & 4.558(2) & 0.7700(7) & 1.5684(9) & 14.403(6) & 0.928(1) & 2.699(1) & 20.642(9) & 0.873(1) & 3.380(2) \\
6 & 8.619(5) & 1.432(1) & 2.241(1) & 29.62(1)& 1.925(2) & 3.810(2) & 48.57(2) & 1.853(3) & 5.045(4) \\
8 & 13.271(8) & 2.183(2) & 2.944(2) & 47.33(2) & 3.258(4) & 4.806(2) & 82.06(4) & 3.154(5) & 6.561(4) \\
10 & 18.26(1) & 3.016(3) & 3.638(2) & 67.19(3) & 4.942(5)& 5.757(3) & 121.79(6) & 4.791(7) & 8.021(5) \\
12 & 23.53(2) & 3.938(4) & 4.309(3) & 88.85(5) & 6.940(8) & 6.637(4) & 168.07(8) & 6.755(9) & 9.441(7) \\
16 & 34.56(3) & 5.977(7) & 5.604(5) & 135.89(7) & 11.90(1) & 8.275(5) & 277.5(1) & 11.70(2) & 12.216(9) \\
20 & 45.95(5) & 8.28(1) & 6.818(7) & 185.9(1) & 18.07(2) & 9.740(6) & 408.2(2) & 17.96(3) & 14.90(1) \\
24 & 57.04(7) & 10.83(2)& 7.911(9) & 237.2(1) & 
25.45(3) & 11.048(7) & 558.9(3) & 25.38(4) & 17.56(1) \\
32 & 77.2(1) & 16.83(2) & 9.66(1) & 336.9(2) & 43.59(4) & 13.233(8) & 912.5(5) & 43.94(6) & 22.68(2) \\
40 & 92.7(2) & 23.64(3) & 10.89(2) & 426.1(3) & 66.16(6) &14.865(9) & 1330.2(7)& 67.30(9) & 27.61(2)\\
48 & 103.0(2) & 31.15(5) & 11.61(2) & 498.4(4) & 91.86(9) &16.08(1) & 1805(1) & 95.3(1) & 32.41(4) \\
64 & 112.3(2) & 45.96(7) & 12.24(2) & 593.2(5) & 150.7(1) &17.46(1) & 2903(2) & 165.7(3) & 41.46(7) \\
80 & 115.3(3) & 58.83(9) & 12.47(2) & 641.4(5) & 212.8(2) &18.07(1) & 4175(5) & 252.4(6) & 50.2(1) \\
96 & 115.7(3) & 69.2(1) & 12.52(2) & 660.6(6) & 271.8(2) & 18.28(1) & 5593(8) & 356(1) & 58.6(2) \\
128 & 115.8(3) & 84.0(2) & 12.55(2) & 670.8(7)& 369.4(3) & 18.40(1) & 8776(12) & 616(2) & 74.1(1) \\
160 & 116.5(3) & 93.5(2) & 12.65(2) & 673.4(6)& 441.3(3) &18.47(2) & 12319(17) & 940(3) & 88.8(2) \\
192 & 115.7(3) & 98.8(2) & 12.64(3) & 673.4(6) & 493.2(4) & 18.47(2) & 16184(23) & 1323(4) & 102.4(3) \\
256 & 115.8(3) & 105.7(2) & 12.62(3) & 674(1) & 559.6(8) & 18.42(3) & 24428(51) & 2285(10) & 126.7(5) \\
320 & 115.8(3) & 109.0(2) & 12.63(3) & 674(1) & 595.6(8) & 18.42(3) & 32864(94) & 3459(18) & 148.6(8) \\
384 & 115.9(3) & 111.1(2) & 12.66(3) & 673(1) & 616.4(8) & 18.48(3) & 41596(166) & 4932(30) &  168(1) \\
512 & 116.0(3) & 113.3(2) & 12.66(3) & 674(1) & 642(1)& 18.45(3) & 57344(507) & 9395(91) & 199(2) \\
\BotRule
\end{tabular}
\caption{Results for $\chi$, $F$ and $\chi(L)$ in the simplest qubit model with no diagonal hops, $U=0.1$ and for an odd number of layers $L_y = 1,3,5$. \label{tab:1}}
\end{table*}

\begin{table*}[!htbp]
\renewcommand{\arraystretch}{1.4}
\setlength{\tabcolsep}{4pt}
\begin{tabular}{c|ccc|ccc|ccc}
\TopRule
 & \multicolumn{3}{c|}{$L_y=2$} &
\multicolumn{3}{c|}{$L_y=4$} & \multicolumn{3}{c}{$L_y=6$} \\ 
\TopRule
$L$ & $\chi$ & $F$ & $\xi(L)$ & $\chi$ & $F$ & $\xi(L)$ & $\chi$ & $F$ & $\xi(L)$ \\
\MidRule
4 & 3.537(2) & 1.065(1) & 1.0771(4) & 13.703(6) & 0.826(1) & 2.803(2) & 20.19(1) & 0.013(1) & 3.481(2) \\
6 & 3.790(2) & 1.766(1) & 1.0700(4) & 32.29(1) & 1.839(2) & 4.082(2) & 53.40(3) & 0.028(3) & 5.396(4) \\
8 & 3.811(2) & 2.284(1) & 1.0669(4) & 55.03(3) & 3.221(4) & 5.258(3) & 94.09(5) & 0.049(5) & 7.114(5) \\
10 & 3.809(2) & 2.655(1) & 1.0669(4) & 80.26(4) & 4.940(6) & 6.343(3) & 142.34(7) & 0.075(8) &  8.773(6) \\
12 & 3.806(2) & 2.917(1) & 1.0671(6) & 107.60(6) & 6.999(9) & 7.325(5) & 198.7(1) & 0.11(2) & 10.45(1) \\
16 & 3.804(2) & 3.243(2) & 1.0660(6) & 166.85(9) & 12.07(1) & 9.178(6) & 332.6(3) & 0.18(3) & 13.65(1) \\
20 & 3.806(2) & 3.425(2) & 1.0661(6) & 231.3(1) & 18.47(2) & 10.849(7) & 494.7(5) & 0.39(6) & 16.77(2) \\
24 & 3.808(2) & 3.534(2) & 1.0668(6) & 299.4(2) & 26.13(3) & 12.388(8) & 684.0(8) & 0.57(9) & 19.87(3) \\
32 & 3.808(2) & 3.649(2) & 1.0669(7) &438.0(3) & 45.06(5) & 15.065(10) & 1138(1) & 1.0(1) & 25.75(3) \\
40 & 3.807(2) & 3.703(2) & 1.0662(7) & 574.6(4) & 68.83(7) & 17.27(1) & 1679(2) & 1.6(2) & 31.58(6) \\
48 & 3.808(2)& 3.735(2) & 1.0663(7) & 701.4(5) & 96.88(10) & 19.10(1) & 2313(4) & 3.2(5) & 37.30(8) \\
64 & 3.810(2) & 3.769(2) & 1.0674(7) & 907.4(8) & 164.4(2) & 21.67(2) & 3800(10) & 10(1) & 48.1(2) \\
80 & 3.807(2) & 3.781(2) & 1.0664(7) & 1044(1) & 242.9(3) & 23.14(2) & 5530(20)& 17(2) &  58.3(2) \\
96 & 3.804(2) & 3.786(2) & 1.0663(7) & 1125(1) & 325.6(3) & 23.94(2) & 7640(30)& 27(4) & 69.5(3) \\
128 & - & - & - & 1191(1) & 486.0(5) & 24.54(2) & 12350(100) & 80(10) & 89.0(7) \\
160 & - & - & - & 1210(1) & 623.2(6) &  24.71(2) & - & - & -  \\
192 & - & - & - & 1213(1) & 732.7(6) & 24.75(2) & - & - & -   \\
256 & - & - & - & 1215(1) & 887.4(8) & 24.76(2) & - & - & -   \\
320 & - & - & - & 1214(1) & 982(1) & 24.76(2) & - & - & - \\
384 & - & - & - & 1214(1) & 1043(1) & 24.76(3) & - & - & -  \\
512 & - & - & - & 1217(1) & 1114(1) & 24.76(3) & - & - & -   \\
\BotRule
\end{tabular}
\caption{Results for $\chi$, $F$ and $\chi(L)$ in the simplest qubit model with no diagonal hops, $U=0.1$ and for an even number of layers $L_y = 2,4,6$. \label{tab:2}}
\end{table*}

\begin{table*}[!htbp]
\renewcommand{\arraystretch}{1.4}
\setlength{\tabcolsep}{4pt}
\begin{tabular}{c|ccc|ccc}
\TopRule
 & \multicolumn{3}{c|}{$J=3$} &
\multicolumn{3}{c}{$J=5$} 
\\ 
\TopRule
$L$ & $\chi$ & $F$ & $\xi(L)$ & $\chi$ & $F$ & $\xi(L)$ \\
\MidRule
4 & 1.9733(3) & 0.05729(7) & 4.089(3) & 0.9761(2) & 0.01535(4) & 5.596(8) \\
6 & 4.5949(7) & 0.2231(1) & 4.427(1) & 2.6657(6) & 0.1231(1) & 4.545(2) \\
8 & 7.631(1) & 0.4112(2) & 5.474(2) & 4.657(1) & 0.2357(2) & 5.658(3) \\
10 & 11.139(2) & 0.6422(3) & 6.542(2) & 6.923(2) & 0.3667(3) & 6.842(3) \\
12 & 15.09(2) & 0.916(4) & 7.61(3) & 9.434(7) & 0.524(1) & 7.97(1) \\
16 & 23.94(4)& 1.615(8) & 9.59(3) & 15.23(1) & 0.897(2) & 10.25(1) \\
20 & 34.10(5) & 2.46(1) & 11.45(3) & 21.91(2) & 1.379(3) & 12.34(2) \\
24 & 45.02(8) & 3.47(1) & 13.17(4) & 29.37(3) & 1.943(5) & 14.39(2) \\
32 & 69.2(1) & 6.01(2) & 16.54(4) & 46.29(4) & 3.348(8) & 18.27(2) \\
40 & 94.7(2) & 9.12(4) & 19.51(6) & 65.08(7) & 5.10(1) & 21.86(3) \\
48 & 120.5(2) & 12.94(5) & 22.06(6) &  84.99(9) & 7.20(2) & 25.14(3) \\
64 & 172.0(4) & 22.23(8) & 26.58(7) & 128.0(1) & 12.29(3) & 31.26(4) \\
80 & 217.7(6) & 33.6(1) & 29.89(8) & 172.7(2) & 18.73(4) & 36.51(5) \\
96 & 256.1(8) & 46.7(2) & 32.28(9) & 216.4(3) & 26.32(6) & 41.06(6) \\
128 & 308(2) & 76.2(4) & 35.2(1) & 297.0(4) & 44.95(9) & 48.25(7) \\
160 & 330(2) & 108.1(6) & 36.4(1) & 361.4(6) & 67.8(1) & 53.00(7) \\
192 & 342(2) & 138(1) & 37.0(1) & 407.3(7) & 92.7(2) & 56.28(8) \\
256 & 348(3) & 189(1) & 37.1(1) & 457(1) & 146.8(4) & 59.2(1) \\
320 & 355(6) & 231(3) & 37.3(1) & 475(1) & 198.2(4) & 60.2(1) \\
384 & 349(7) & 251(4) & 37.3(1) & 481(2) & 242.7(6) & 60.6(1) \\
512 & 350(9) & 284(6) & 36.9(2) & 485(2) & 311.3(8) & 60.8(2) \\
\BotRule
\end{tabular}
\caption{Results for $\chi$, $F$ and $\chi(L)$ in the extended qubit model with diagonal hops but only one layer ($L_y=1$). The data are for two values of $J=3$ and $J=5$ at $U=0$.\label{tab:3}}
\end{table*}

\end{document}